\documentclass[twocolumn,showpacs,aps,prc,superscriptaddress,floatfix]{revtex4}

\usepackage[dvips]{graphicx}

\begin{document}

\title{
Measurement of the Generalized Polarizabilities of
the Proton in Virtual Compton Scattering at 
$Q^2$=0.92 and 1.76 GeV$^2$ :
II. Dispersion Relation Analysis }

\author{G.~Laveissi\`{e}re}
\affiliation{Universit\'{e} Blaise Pascal/IN2P3, F-63177 Aubi\`{e}re, France}
\author{L.~Todor}
\affiliation{Old Dominion University, Norfolk, VA 23529}
\author{N.~Degrande}
\affiliation{University of Gent, B-9000 Gent, Belgium}
\author{S.~Jaminion}
\affiliation{Universit\'{e} Blaise Pascal/IN2P3, F-63177 Aubi\`{e}re, France}
\author{C.~Jutier}
\affiliation{Universit\'{e} Blaise Pascal/IN2P3, F-63177 Aubi\`{e}re, France}
\affiliation{Old Dominion University, Norfolk, VA 23529}
\author{R.~Di Salvo}
\affiliation{Universit\'{e} Blaise Pascal/IN2P3, F-63177 Aubi\`{e}re, France}
\author{L.~Van Hoorebeke}
\affiliation{University of Gent, B-9000 Gent, Belgium}
\author{L.C.~Alexa}
\affiliation{University of Regina, Regina, SK S4S OA2, Canada}
\author{B.D.~Anderson}
\affiliation{Kent State University, Kent OH 44242}
\author{K.A.~Aniol}
\affiliation{California State University, Los Angeles, CA 90032}
% ...
\collaboration{The Jefferson Lab Hall A Collaboration}
\noaffiliation


\begin{abstract}
Virtual Compton Scattering is studied at 
the Thomas Jefferson National Accelerator Facility
in the energy domain below pion threshold and 
in the $\Delta(1232)$ resonance region.
The data analysis is based on the Dispersion Relation (DR) approach.
The electric and magnetic Generalized Polarizabilities (GPs) of the proton
and the structure functions 
$P_{LL}-P_{TT}/\epsilon$ and $P_{LT}$
are determined at four-momentum transfer squared
$Q^2$= 0.92 and 1.76 GeV$^2$. The DR analysis is consistent with
the low-energy expansion analysis. The world data set indicates that
neither the electric nor magnetic GP follows a simple dipole form.
\end{abstract}


\maketitle

%
%
%
Virtual Compton Scattering (VCS) $\gamma^* p \to \gamma p$
has developed in the last decade as a powerful tool to
study the nucleon structure.
At low CM-frame energy $W$, the VCS amplitude is parametrized
as a function of the Generalized Polarizabilities
(GPs) of the proton~\cite{Guichon:1995pu}, which depend on
the four-momentum transfer squared $Q^2$ of the
virtual photon. 
These new observables have become the subject of 
experimental investigation via the study of 
photon electroproduction $e p \to ep \gamma$.
In a companion Letter (referred to as I) 
we present an extraction of the structure functions
$P_{LL}-P_{TT}/\epsilon$ and $P_{LT}$
from data below $N \pi$ threshold~\cite{Jaminion:2003le},
following the low-energy expansion (LEX) 
formalism~\cite{Guichon:1995pu}.

%
B. Pasquini \textit{et al.} recently developed a formalism 
for VCS  based
on Dispersion Relations (DRs)~\cite{Pasquini:2001yy}.  
In this Letter we report
a determination of the electric and magnetic GPs of 
the proton 
($\alpha_E(Q^2)$ and $\beta_M(Q^2)$, respectively) 
from an analysis based on the DR model.
The structure functions 
$P_{LL}-P_{TT}/\epsilon$ and $P_{LT}$ 
are also extracted. The method uses
the photon electroproduction cross section
measured in the E93-050 experiment~\cite{Bertin:1993} 
at Jefferson Lab (JLab).
 
In the DR formalism, the VCS amplitude is predicted from the
MAID parametrization~\cite{Drechsel:1998hk} 
of pion electroproduction,
$\pi^0$ and $\sigma$-meson $t$-channel exchange, plus two
other $Q^2$-dependent functions (subtraction constants).
The latter are unconstrained phenomenological contributions to
$\alpha_E(Q^2)$ and $\beta_M(Q^2)$;
therefore these GPs are not fixed by the DR model. 
They are written as:
\begin{eqnarray}
\alpha_E(Q^2) \ = \ \alpha_E^{\pi N}(Q^2) \ + \ 
{ \displaystyle [ \ \alpha_{E}^{exp}   -  
\alpha_{E}^{\pi N} \ ]  _{Q^2=0} 
\over
\displaystyle (\ 1 + Q^2/  \Lambda_{\alpha}^2 \ )^2 }
\label{eq01}
\end{eqnarray}
(same relation for $\beta_M$ with parameter $\Lambda _{\beta}$) 
where $\alpha_E^{\pi N}$ ($\beta_M^{\pi N}$)
is the $\pi N$ dispersive contribution evaluated using the 
MAID analysis, and $\alpha_E^{exp}$ ($\beta_M^{exp}$) is 
the experimental value at $Q^2=0$. 
The mass coefficients 
$\Lambda _{\alpha}$ and $\Lambda _{\beta}$ 
are free parameters to be fitted experimentally.
We note that the choice of a 
dipole form in Eq.~\ref{eq01} is not compulsory.
A more fundamental property of the DR model is that,
up to the $N \pi \pi$ threshold, it provides 
a rigorous treatment of the higher order terms
in the VCS amplitude, beyond 
the lowest order GPs given by the LEX~\cite{Guichon:1995pu}.
This feature allows the inclusion of data in the 
$\Delta(1232)$ resonance region 
in an extraction of GPs based on the DR approach.

\begin{figure}[t]
\includegraphics[width=8.4cm]{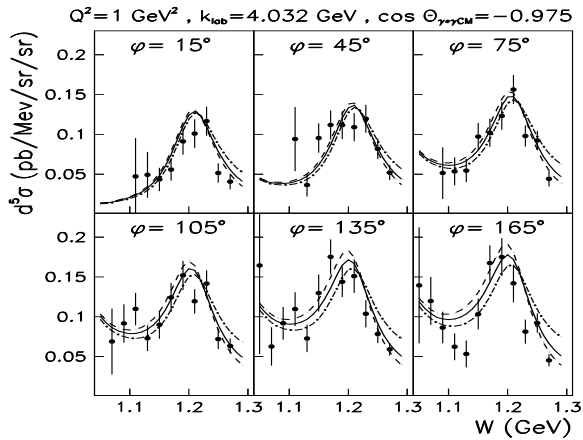}
\caption{\label{secdelta}
$(ep \to ep \gamma)$ cross section for data set I-b
in six intervals of the azimuthal angle $\varphi$ 
(angle between lepton and hadron planes) as a function of $W$.
By symmetry the statistics for $\varphi$= 
180$^{\circ}$ to 360$^{\circ}$ are also included.
Only statistical errors are shown. 
The solid curve is the prediction of the DR model for
parameter values
($\Lambda _{\alpha}$, $\Lambda _{\beta}$)=(0.7, 0.6) GeV,
while the dashed curve is for (0.5, 0.4) GeV and
the dash-dotted curve for (0.9, 0.8) GeV, respectively.
}
\end{figure}

\begin{figure}[h]
\includegraphics[width=8.5cm]{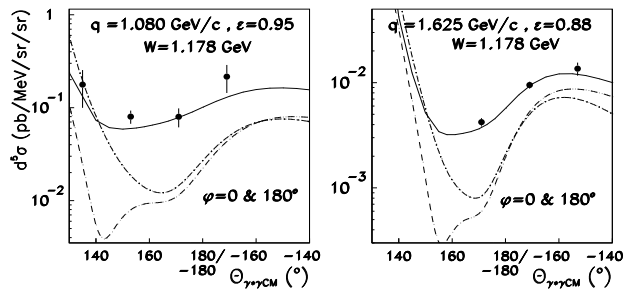}
\caption{\label{secda1da2}
$(ep \to ep \gamma)$ cross section for data sets I-a (left) 
and II (right). 
The outgoing photon is emitted
in the lepton plane ($\varphi=0^{\circ}$ or $180^{\circ}$). 
The abscissa is 
the polar angle  $\theta_{\gamma ^* \gamma CM}$, 
with negative values corresponding to 
$\varphi=180^{\circ}$ as in Ref.~\cite{Roche:2000ng}.
Only statistical errors are shown.
The full curve is the DR model prediction using the
values of ($\Lambda _{\alpha}$, $\Lambda _{\beta}$) fitted
to each of the two data sets (see Table~\ref{resultslalb}).
The dashed (dash-dotted) curve is 
the BH+Born (plus a first-order GP effect) cross section.
 }
\end{figure}

%
%
The JLab experiment 
uses the 4 GeV Continuous Electron Beam
Accelerator and the Hall A 
instrumentation~\cite{Alcorn:2003}. More details can be found 
elsewhere~\cite{Laveissiere:2003pi,
Degrande:2001th,Jaminion:2001th,Jutier:2001th,Laveissiere:2001th,
Todor:2000th}.
The present study involves all the $(ep \to ep \gamma)$ events 
with $W<1.28$ GeV.
The events are divided into three independent subsets listed in 
Table~\ref{datasets}.
For data sets I-a and II, by far most of the events 
lie below pion threshold; actually these two sets have 
also been used in the LEX analysis~\cite{Jaminion:2003le} 
considering only events such that $W < (M_p+M_{\pi})$.
Data set I-b covers mainly the $\Delta(1232)$ 
resonance region in $W$.
\begin{table}[h]
\caption{\label{datasets} Data sets for the DR analyses.
 }
\begin{ruledtabular}
\begin{tabular}{ccc} 
data set & $Q^2$-range (GeV$^2$) & $W$-range   \\
\hline
I-a & [0.85, 1.15] & mostly $< \pi N$ threshold \\
I-b & [0.85, 1.15] & mostly $\Delta$ resonance \\
II  & [1.60, 2.10] & mostly $< \pi N$ threshold \\
\end{tabular}
\end{ruledtabular}
\end{table}
%
The three data sets of Table~\ref{datasets} 
are the subject of three distinct DR analyses. 
The basic ingredients for the cross-section determination
are common to all our analyses of this 
experiment~\cite{Laveissiere:2003pi,Degrande:2001th,
Jaminion:2001th,Jutier:2001th,Laveissiere:2001th,Todor:2000th}.
They include a dedicated Monte-Carlo 
simulation~\cite{VanHoorebeke} to obtain the acceptance,
proper cuts to eliminate background, 
the application of radiative corrections~\cite{Vanderhaeghen:2000ws},
the calibration of experimental offsets and 
the absolute normalization of the experiment.
It is important to have a realistic shape for the sampling
cross section in the Monte-Carlo in order to calculate
an accurate solid angle. 
For this purpose, simulated events are sampled in
the DR model cross section ($d^5 \sigma^{DR}$)
which reproduces the enhancement of
the $\Delta$ resonance.
$d^5 \sigma^{DR}$ depends on two free parameters 
$\Lambda _{\alpha}$ and $\Lambda _{\beta}$ (cf. Eq.~\ref{eq01})
which are iteratively fitted by a $\chi^2$ minimization at
the cross section level.
%
%
When $W$ increases, the acceptance is reduced 
to backward polar angles $\theta_{\gamma ^* \gamma CM}$
(angle between the virtual and the final photons in the
$(\gamma p)$ CM frame).
Cross-section data obtained in this angular region are 
shown in Figs.~\ref{secdelta} and \ref{secda1da2}.
Kinematical conditions are defined on the plots, 
$k_{lab}$ being the incoming beam energy (Fig.~\ref{secdelta}), 
$q$ the virtual photon CM momentum and $\epsilon$ the
virtual photon polarization (Fig.~\ref{secda1da2}).
Figure~\ref{secdelta} clearly shows the 
$\Delta$ resonance excitation
and the various curves indicate the sensitivity of the
DR model to its free parameters.
In contrast with Fig.~\ref{secdelta}, Fig.~\ref{secda1da2} 
shows how the Bethe-Heitler+Born calculation, or the 
addition of a first-order GP effect 
as in the LEX approach~\cite{Jaminion:2003le},
fails to reproduce the measured cross section above pion threshold.


\begin{table}[b]
\caption{\label{resultslalb} 
Upper part: 
dipole mass parameters $\Lambda_{\alpha}$ and $\Lambda_{\beta}$
obtained by fitting the three data sets
independently. Lower part: electric and magnetic GPs evaluated
at $Q^2$= 0.92 GeV$^2$ (data sets I-a, I-b) and 1.76 GeV$^2$
(data set II).
The first and second errors are 
statistical and total systematic errors, respectively.
 }
\begin{ruledtabular}
\begin{tabular}{ccc}
 data set
& $\Lambda_{\alpha}$ (GeV) & $\Lambda_{\beta}$  (GeV) \\
\hline
I-a &  0.741 $\pm$ 0.040 $\pm$ 0.175 & 0.788  $\pm$ 0.041 $\pm$ 0.114  \\ 
I-b & 0.702  $\pm$ 0.035 $\pm$ 0.037 & 0.632  $\pm$ 0.036 $\pm$ 0.023  \\
II &  0.774 $\pm$ 0.050 $\pm$ 0.149 & 0.698  $\pm$ 0.042 $\pm$ 0.077   \\
\hline
data set   & $\alpha_E(Q^2)$  ($10^{-4}$fm$^3$) 
& $\beta_M(Q^2)$  ($10^{-4}$fm$^3$) \\
 \hline
I-a         &  1.02  $\pm$  0.18  $\pm$  0.77    
            & 0.13  $\pm$  0.15  $\pm$  0.42  \\ 
I-b         & 0.85  $\pm$  0.15  $\pm$  0.16 
            & 0.66  $\pm$  0.11  $\pm$  0.07   \\
II          & 0.52  $\pm$  0.12  $\pm$ 0.35
            & 0.10 $\pm$  0.07 $\pm$ 0.12   \\
\end{tabular}
\end{ruledtabular}
\end{table}


The results for $\Lambda _{\alpha}$ and $\Lambda _{\beta}$
are presented in Table~\ref{resultslalb}.
Systematic errors are calculated from the
same four uncertainties as in the LEX analysis~\cite{Jaminion:2003le}.
The resulting error bars differ from one data set to 
another; this comes from the various data sets having a different
phase space coverage, and a different sensitivity to both the
physics and the sources of systematic errors.
The reasonably good $\chi^2$ of the fits (1.3 to 1.5) 
indicates that 
the DR model works well in our kinematics and allows 
a reliable extraction of GPs, both below 
and above pion threshold.
The compatibility between the three fitted values of 
$\Lambda_{\alpha}$ within errors 
(same for $\Lambda_{\beta}$) suggests that the dipole form of 
Eq.~\ref{eq01} is rather realistic,
at least in the range $Q^2$ $\sim$ 1-2 GeV$^2$.
We also note the close agreement between the obtained values of
$\Lambda _{\alpha}$ and  $\Lambda _{\beta}$.

\begin{figure}[t]
\includegraphics[width=8.5cm]{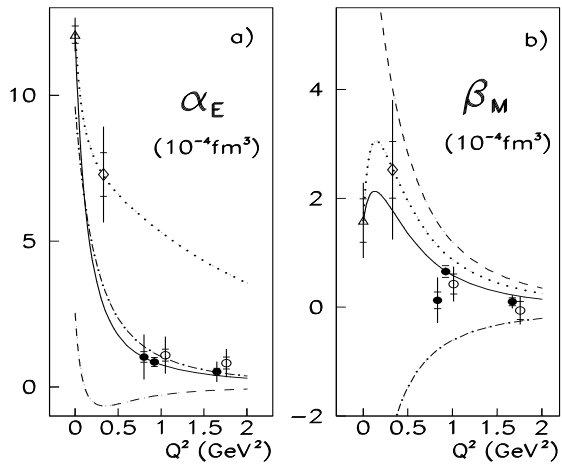}
\caption{\label{plotsf}
Compilation of the data on electric (a) and
magnetic (b) GPs. 
Data points are from Refs.~\cite{OlmosdeLeon:2001zn}
({\tiny $\bigtriangleup$}), the LEX analyses of 
MAMI~\cite{Roche:2000ng} ($\diamond$) and 
JLab~\cite{Jaminion:2003le} ($\circ$) and the present 
DR results ($\bullet$).
Some JLab points are shifted in abscissa for better visibility.
The inner error bar is statistical; the outer one is the total
error (quadratic sum of statistical and systematic errors). 
The curves show the results of calculations in the DR 
model (see text).
}
\end{figure}

In the DR model these results directly translate into values for
the electric and magnetic GPs, using Eq.~\ref{eq01}. 
Table~\ref{resultslalb} gives the result 
of this evaluation of
 $\alpha_E(Q^2)$ and $\beta_M(Q^2)$ at 
$Q^2$= 0.92 GeV$^2$ (data sets I-a and I-b) and
$Q^2$= 1.76 GeV$^2$ (data set II).
These points are shown in Fig.~\ref{plotsf} together
with the point at $Q^2$= 0~\cite{OlmosdeLeon:2001zn}
and the points derived from the LEX 
analyses~\cite{Roche:2000ng,Jaminion:2003le}.
It must be noted that the latter do not directly yield the GPs,
but the structure functions $P_{LL} -P_{TT}/\epsilon$ and $P_{LT}$,
which are combinations of GPs. Therefore to obtain the ``LEX points''
of Fig.~\ref{plotsf} we have extracted
 $\alpha_E(Q^2)$ and $\beta_M(Q^2)$ from the 
measured structure functions~\cite{Roche:2000ng,Jaminion:2003le}
according to the formulas~\cite{Pasquini:2001yy}:
\begin{eqnarray}
P_{LL}-{1 \over \epsilon} P_{TT}  =  
{ 4 M_p \over \alpha_{_{\rm QED}} } 
G_E^p  \ \alpha_E (Q^2) + \mbox{ [spin-flip GPs] } \label{eq02a} \\
P_{LT} =  - { 2 M_p \over \alpha_{_{\rm QED}} } 
 \sqrt{ {q^2 \over Q^2} } 
G_E^p  \ \beta_M (Q^2) + \mbox{ [spin-flip GPs] }  \label{eq02b}
\end{eqnarray}
where $\alpha_{_{\rm QED}}$ is the fine-structure constant and  
$G_E^p$ is the proton electric form factor.
In this extraction the spin-flip GP terms are
predicted by the DR model (so the result is
model-dependent) and the parametrization of 
$G_E^p$ is taken from Ref.~\cite{Brash:2001qq}.
The solid curve on Fig.~\ref{plotsf} is the full DR calculation,
split into its dispersive $\pi N$ contribution (dashed curve) and
the remaining asymptotic contribution
(dash-dotted curve, dipole term of Eq.~\ref{eq01}) for
$\Lambda_{\alpha}$=0.70 GeV and $\Lambda_{\beta}$=0.63 GeV,
as fitted on the JLab data set I-b.
The $\pi N$ intermediate states 
contribute very little to the electric polarizability 
at finite $Q^2$ (Fig.~\ref{plotsf}-a)
and they create
a small dia-electric effect ($\alpha_E^{\pi N}<0$). 
The $\pi N$ contribution to the magnetic polarizability in 
Fig.~\ref{plotsf}-b is strongly paramagnetic,
predominantly arising from the $\Delta(1232)$ resonance. 
In the DR formalism, 
this is cancelled by a strong diamagnetic term originating from
$\sigma$-meson $t$-channel exchange, and 
parametrized by $\Lambda_{\beta}$.
The dotted curve is the full DR calculation evaluated for
$\Lambda_{\alpha}$=1.79 GeV and $\Lambda_{\beta}$=0.51 GeV,
which reproduces the MAMI LEX data.
The fact that there is no unique value of
($\Lambda_{\alpha}$, $\Lambda_{\beta}$) agreeing with
all data points suggests that the dipole form of Eq.~\ref{eq01},
although working 
well in the range 1-2 GeV$^2$ as mentioned above,
is not valid over the entire range of $Q^2$. This further
suggests that the global behavior
of the GP  $\alpha_E(Q^2)$ does not follow a simple dipole form.
%
\begin{table}[b]
\caption{\label{resultssf}
VCS structure functions obtained by the DR analyses
of the three data sets. 
The first and second errors are 
statistical  and total systematic errors, respectively.
}
\begin{ruledtabular}
\begin{tabular}{ccccc}
data  & $Q^2$ & $\epsilon$ &
$P_{LL}-P_{TT}/\epsilon$ & $P_{LT}$ \\
set & (GeV$^2$) & \ & (GeV$^{-2}$) & (GeV$^{-2}$) \\
 \hline
I-a &  0.92  & 0.95 & \textbf{1.70} & \textbf{-0.36} \\
\ &  \ & \ 
& { $\pm$ 0.21 $\pm$ 0.89 }
& { $\pm$ 0.10 $\pm$ 0.27 } \\
I-b  & 0.92  & 0.95 & \textbf{1.50} & \textbf{-0.71} \\
\ &  \ & \ 
& { $\pm$ 0.18 $\pm$ 0.19 }
& { $\pm$ 0.07 $\pm$ 0.05 } \\
II &  1.76 & 0.88 & \textbf{0.40} & \textbf{-0.087} \\
\ &  \ & \
& { $\pm$ 0.05 $\pm$ 0.16 }
& { $\pm$ 0.019 $\pm$ 0.034 } \\
\end{tabular}
\end{ruledtabular}
\end{table}

%
Finally, the results of our DR analyses 
can be expressed
in terms of VCS structure functions. 
Using the GP values of Table~\ref{resultslalb}
and the DR model prediction for $P_{TT}$, the structure
functions $P_{LL}-P_{TT}/\epsilon$ and $P_{LT}$
are determined from Eqs.~\ref{eq02a} and~\ref{eq02b}.
Their values are reported in Table~\ref{resultssf}.
To facilitate the comparison with the LEX results,
the present determination is made at similar kinematics
in $Q^2$ and $\epsilon$ (see Table~\ref{resultssf}).
The agreement with the values obtained in 
the LEX analysis~\cite{Jaminion:2003le} is very satisfactory. 
The results from the DR analysis of the (I-b) data set, 
obtained in the $\Delta$ region, yields
smaller error bars due to an enhanced sensitivity
to the GPs.


In summary we have analyzed  
the process $ep \to ep \gamma$  at JLab
using a Dispersion Relation approach.
The VCS structure functions obtained in this 
approach are in good agreement with the ones extracted 
using the low-energy expansion.
We performed the first determination of GPs by analyzing data
in the $\Delta(1232)$ resonance region. 
This opens up new possibilities 
to extract GPs from experiments, especially at higher $Q^2$. 

We thank the JLab accelerator staff and the Hall A technical staff
for their dedication.
This work was supported by DOE contract DE-AC05-84ER40150 under
which the Southeastern Universities Research Association (SURA)
operates the Thomas Jefferson National Accelerator Facility. We
acknowledge additional grants from the US DOE and NSF, the French
Centre National de la Recherche Scientifique and Commissariat \`a
l'Energie Atomique, the Conseil R\'egional d'Auvergne, the
FWO-Flanders (Belgium) and the BOF-Gent University. 
We thank for the hospitality of ECT* (Trento) during VCS
workshops where this work was discussed.

\bibliography{/users/divers/cebaf/vcs/publications/struc_func/common}

\end{document}